\documentclass{PoS}
\usepackage{bbold}
\usepackage{graphicx}
\def \bea{\begin{eqnarray}}
\def \beq{\begin{equation}}
\def \eea{\end{eqnarray}}
\def \eeq{\end{equation}}
\def \ob{\overline{B}}

\title{$B_s$ Decays and Mixing}

\ShortTitle{$B_s$ Decays and Mixing}

\author{\speaker{Jonathan L. Rosner}\\
        Enrico Fermi Institute and Department of Physics\\
        University of Chicago\\
        E-mail: \email{rosner@hep.uchicago.edu}}

\author{Michael Gronau\\
        Department of Physics, Technion -- Israel Institute of Technology\\
        Haifa, Israel\\
        E-mail: \email{gronau@physics.technion.ac.il}}

\abstract{Theoretical remarks are offered regarding recent hadron collider
results on the mixing and decays of $B_s$ mesons.  Topics covered include:
(1) CP-violating mixing in $B_s(\ob_s) \to J/\psi \phi$,
(2) the D0 dimuon charge asymmetry, (3) information from triple products,
(4) $B_s \to J/\psi f_0$, (5) new physics constraints, (6) some illustrative
new physics scenarios.}

\FullConference{The 13th International Conference on B-Physics at Hadron Machines - Beauty2011,\\
		April 04-08, 2011\\
		Amsterdam, The Netherlands}

\begin{document}

\section{Introduction}

Recent results on $B_s$ decays and mixing have been presented by the CDF and D0
Collaborations at the Fermilab Tevatron and the LHCb Collaboration at CERN.  We
begin by discussing CP-violating mixing in $B_s~(\ob_s) \to J/\psi \phi$.
Experiments at CDF and D0 suggested a mixing phase $\beta_s$ much larger than
that in the Standard Model (SM).  With such a large phase, we pointed out that
time-dependent decays should display explicit time-dependence \cite{GRtd}.  We
update that analysis in Section 2.

The D0 Collaboration has presented evidence for a charge asymmetry in same-sign
dimuons produced in $\bar p p$ collisions at $\sqrt{s} = 1.96$ TeV
\cite{D0mumu}.  We suggest in Section 3 a test of whether this asymmetry is due
to decays of $b$ quarks, as claimed, or background sources such as kaons
\cite{GRmumu}.
 
In Section 4 we discuss what triple products in $B_{(s)} \to V_1 V_2$
actually measure.  The answer \cite{Datta} is CP violation, but only under
certain conditions.  The study of $B_s \to J/\psi f_0$, mentioned in Section
5, avoids the angular analysis needed to interpret $B_s \to J/\psi \phi$.
In Section 6, we note constraints on new physics, and comment in Section 7 on a
couple of scenarios for consideration should any hints for physics beyond the
SM be borne out by further tests.  We conclude in Section 8.

\section{CP violation in interference between $B_s$--$\ob_s$ mixing and
$B_s \to J/\psi \phi$ decay}

For formalism we refer to \cite{LNCKMf}.  $B_s$--$\ob_s$ mixing is expected
to be dominated by the top quark in box graphs.  The observed values $\Delta
m_s=(17.77 \pm 0.10 \pm 0.07)$ ps$^{-1}$ (CDF \cite{Abulencia:2006}) and
$(17.63 \pm 0.11 \pm 0.04)$ ps$^{-1}$ (LHCb \cite{LHCbDeltam}) agree with SM
predictions.  Denoting
\beq
|B_{sL} \rangle = p |B_s \rangle + q | \bar B_s \rangle~;~~|B_{sH} \rangle=p
|B_s \rangle-q|\bar B_s \rangle~,
\eeq
we expect for $\Delta \Gamma \ll \Delta m$, $q/p \simeq \exp(2 i \beta_s)$,
$\beta^{\rm SM}_s = - {\rm Arg} (-V_{ts}^*V_{tb}/V_{cs}^*V_{cb}) = (1.04 \pm
0.05)^\circ$ \cite{LNCKMf}.  The SM $B_s \to J/\psi \phi$ CP asymmetry then
should be governed by the small mixing phase $\phi_M = -2 \beta^{\rm SM}_s$.

In 2008, CDF \cite{CDFmix08} and D0 \cite{D0mix08} favored a mixing phase
differing from $-2 \beta_s$ by $\sim
2.2 \sigma$ based on the decay $B_s \to J/\psi \phi$.  At that time we pointed
out that such a large mixing phase (the illustrative value was then $\phi_M = -
44^\circ$ \cite{CDFmix08}) would imply detectable time-dependence of angular
distribution coefficients, differing for tagged $B_s$ and $\ob_s$ \cite{GRtd}.

\begin{figure}
\begin{center}
\includegraphics[width = 0.75\textwidth]{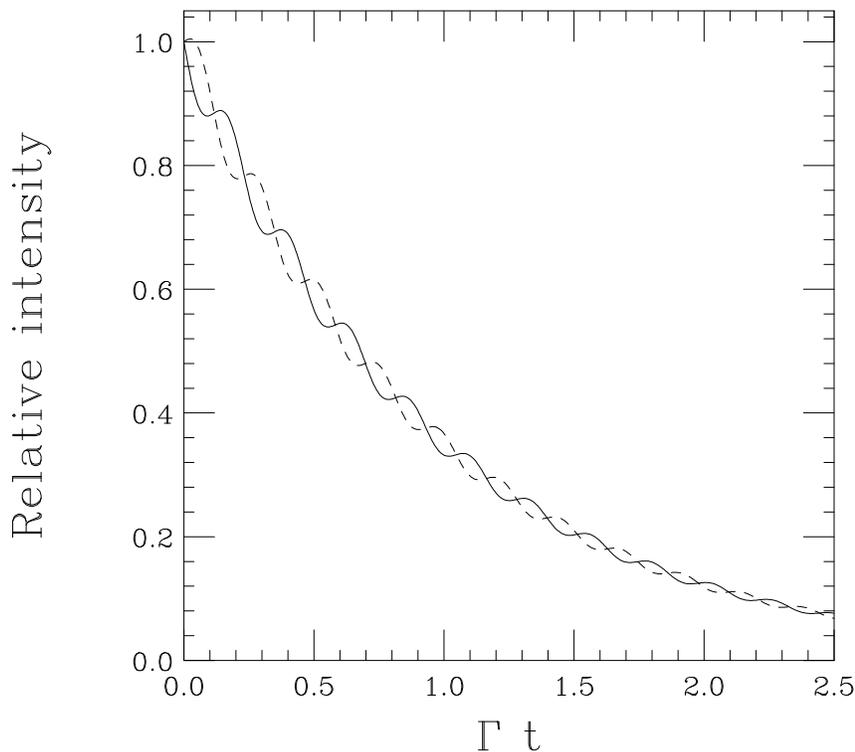}
\end{center}
\caption{Relative intensities of ${\cal T}_+$ signals as functions of
$\Gamma t$, for $B_s$ tags (solid) and $\ob_s$ tags (dashed).  This figure
represents an update of a similar one in Ref.\ \cite{GRtd}.
\label{fig:oscs}}
\end{figure}

We review the discussion briefly.  For a CP test, one tags the flavor at $t=0$,
denoting $\eta = \pm 1$ for a tagged $(B_s,\bar B_s)$.  The coefficients of
helicity amplitudes $|A_\parallel|^2$, $|A_\perp|^2$ describing different
angular dependences are denoted by ${\cal T}_+$, ${\cal T}_-$, where 
\beq
{\cal T}_{\pm}\equiv e^{- \Gamma t} [\cosh(\Delta \Gamma t)/2 \mp \cos(\phi_M)
\sinh(\Delta \Gamma t)/2 \pm \eta \sin(\phi_M) \sin (\Delta m_s t)]~.
\eeq
Taking $\phi_M = -44^\circ,~\Delta \Gamma/\Gamma = 0.228$, and assuming the
tagging $\eta$ to be diluted by a factor of 0.11, we concluded that wiggles
should be distinguishable between the $B_s$-tagged and $\ob_s$-tagged
${\cal T}_\pm$ distributions.  We advocated making such a plot as evidence for
CP violation in $B_s \to J/\psi \phi$ at a level beyond the SM.  Here we update
our estimate of $t$-dependence, finding the
oscillations a bit smaller, but still visible.  We take
$\phi_M = (-39\pm17)^\circ$ based on an average between CDF \cite{CDF10206,%
CDFphim} and D0 \cite{D0phim} values, choose $\Delta \Gamma/\Gamma = 0.143$
based on an average between CDF ($0.075 \pm 0.035 \pm 0.010$) and D0 ($0.15
\pm 0.06 \pm 0.01$), and continue to assume a dilution factor of 11\%.  The
resulting plot is shown in Fig.\ \ref{fig:oscs}.

At this Conference, LHCb presented data restricting $\phi_M$ to the range
$[-2.7,-0.5]$ \cite{LHCbphim} (68\% c.l.), $1.2\sigma$ from the SM.
We are eagerly awaiting data from ATLAS and CMS.

\section{D0 dimuon asymmetry -- Is it due to $b$'s?  $K$'s?}

The SM predicts a small asymmetry in the yield of same-sign muon
pairs due to $b \bar b$ production followed by meson $\Leftrightarrow$
antimeson oscillation: $A_{sl}^b \equiv \frac{N^{++} - N^{--}}{N^{++} + N^{--}}
= (-2.0 \pm 0.3)\times 10^{-4}$ \cite{LN2011}.  The D0 Collaboration reports a
much larger value, $A_{sl}^b = (-9.57 \pm 2.51 \pm 1.46) \times 10^{-3}$,
nearly 50 times the SM value \cite{D0mumu}.  (CDF is not ready to report such
a measurement but has quoted a new average mixing parameter $\bar \chi$
\cite{Harr}.)

D0 has interpreted its result as $3.2 \sigma$ evidence for CP violation in
neutral $B$ mixing.  They have performed 16 systematic checks for which their
results are found consistent with their nominal ones.  Estimating the correct 
kaon decay backgrounds is crucial.

We have suggested a test \cite{GRmumu} to see if a smaller asymmetry is
obtained in a sample depleted in $b \bar b$ pairs.  If one reduces the
maximum allowed impact parameter of muon tracks, the signal should vanish more
rapidly than background.  The effect of our suggestion, an impact parameter
cut of $b<100 \mu$m, is not yet known to us.

\begin{figure}
\begin{center}
\includegraphics[width=0.75\textwidth]{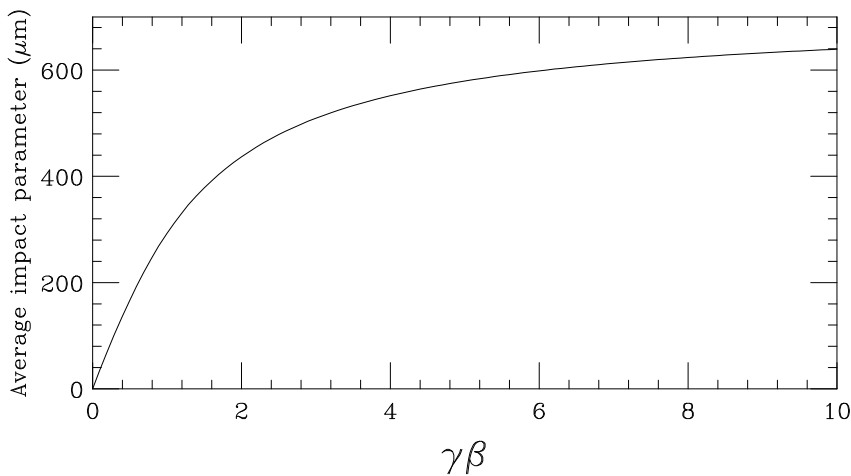}
\end{center}
\caption{Dependence of $\langle b \rangle$ on $\gamma \beta$ \cite{GRmumu}.
\label{fig:bgb}}
\end{figure}

We denote quantities in the $B$ rest frame with an asterisk (*) and those in
the lab frame with none.  The lab energy of the $B$ is $E_B = \gamma m_B =
m_B/\sqrt{1-\beta^2}$.  Muon angles with respect to the $B$ boost are denoted
by $\theta^*$ in the $B$ rest frame and $\theta$ in the lab. The transformation
between them is $\sin \theta=\sin \theta^*/[\gamma(1 + \beta \cos \theta^*)]$.
The isotropy of muon emission in $\cos \theta^*$ can be used to calculate
the average values of $\sin \theta$ and $b = \gamma \beta \sin \theta c \tau$,
where $c \tau = 450 \mu$m and
\beq
\langle \sin \theta \rangle = \frac{1}{2} \int_0^\pi \frac{\sin^2 \theta^* d
\theta^*}{\gamma(1 + \beta \cos \theta^*)} = \frac{\pi} {2} \frac{1}{1 +
\gamma}~.
\eeq
The dependence of $\langle b \rangle$ on $\gamma \beta$ is shown in Fig.\
\ref{fig:bgb}.

An eyeball fit to the CDF $b$ distribution \cite{CDFb} gives $\langle b
\rangle = 350~\mu$m.  Table \ref{tab:cut} denotes the effect of discarding
events with $b$ exceeding various values of $b_0$.

\begin{table}
\caption{Fraction of events remaining for a given $\langle b \rangle$ when
events with $b > b_0$ are discarded \cite{GRmumu}.
\label{tab:cut}}
\begin{center}
\begin{tabular}{l c c c c c} \hline \hline
\qquad $b_0$ ($\mu$m) & 100 & 200 & 300 & 400 & 500 \\
$\langle b \rangle$ ($\mu$m) & & & & & \\ \hline
150 & 0.237 & 0.542 & 0.748 & 0.866 & 0.930 \\
300 & 0.080 & 0.237 & 0.400 & 0.542 & 0.658 \\
450 & 0.040 & 0.129 & 0.237 & 0.347 & 0.450\\ \hline \hline
\end{tabular}
\end{center}
\end{table}

The D0 Collaboration defines a transverse impact parameter $b_\perp$ relative
to the closest primary vertex and a longitudinal distance $b_\parallel$ from
the point of closest approach to this vertex.  They choose $b_\perp<3000~\mu$m
and $b_\parallel < 5000~\mu$m.  These are related to $b$ as follows.  The
transverse and longitudinal components of muon momentum in the lab are
$p^\mu_\perp = p^\mu \sin \psi$, $p^\mu_\parallel = p^\mu \cos \psi$.  The
distance $d$ of a point along the $\mu$ trajectory from the vertex is
$d^2 = b_\perp^2 + (s \sin \psi)^2 + (s \cos \psi - b_\parallel)^2$, where
$s=$ is the distance along the $\mu$ trajectory from the transverse point of
closest approach.  The minimum of $d$ is $b = d_{\rm min} = [b_\perp^2 +
(b_\parallel \sin \psi)^2]^{1/2}$.  Little signal reduction is seen with
$b_\perp < 500~\mu$m, $b_\parallel < 500~\mu$m \cite{D0mumu}, but we advocate
a tighter cut.  The key question remains with regard to D0 muons:  are they
really from $b$ decays?
This question should be answered by imposing an upper bound of $b_0 < 100
\mu$m on the impact parameter $b_0$.

\section{What do triple products in $B_{(s)} \to V_1 V_2$ measure?}

A spinless particle decaying to four spinless particles gives rise to three
independent momenta in its rest frame.  One can form a T-odd expectation value
out of (e.g.) $\mathbf{p_1 \times p_2 \cdot p_3}$ \cite{Datta,Valencia:1989}.
A famous
example is the asymmetry of $(13.6 \pm 1.4 \pm 1.5)\%$ in $K_L \to \pi^+
\pi^- e^+ e^-$ reported by the KTeV Collaboration \cite{KTeV}.  However,
what if two or more of the final-state particles are identical?

Consider the double-Dalitz decay of a CP-mixture (like $K_L$) to $e^+ e^- e^+
e^-$.  (see, e.g., \cite{NA31}).  For low $M(e^+ e^-)$ this process is like
$K_L \to \gamma \gamma$, with photons having relative linear polarizations
$(\parallel,\perp)$ for CP = $(+,-)$.  Interference between CP-even and -odd
decays can give a non-vanishing value of $\langle \sin \phi \cos \phi \rangle$,
where $\phi$ is the angle between normals to the $e^+e^-$ planes.

Now consider the case of $B \to V_1 V_2$, with each $V$ decaying to two
pseudoscalar mesons $P$.  (For an extensive discussion of the formalism, see
\cite{DL}.)  One extracts triple products (TPs) from angular analyses:
\beq
A_T \equiv \frac{\Gamma({\rm TP} > 0) - \Gamma({\rm TP} < 0)}
                {\Gamma({\rm TP} > 0) + \Gamma({\rm TP} < 0)}~;
~~{\rm TP} \equiv p_1 \cdot (p_2 \times p_3)~;
\eeq
they are tiny in the SM.  A true T-violation is signified by
\beq
{\cal A}_T^{\rm true} \equiv
\frac{\Gamma({\rm TP} > 0) + \bar
\Gamma({\rm TP} > 0) - \Gamma({\rm TP} < 0) - \bar \Gamma({\rm TP} < 0)}
{\Gamma({\rm TP} > 0) + \bar \Gamma({\rm TP} > 0) + \Gamma({\rm TP} < 0)
+ \bar \Gamma({\rm TP} < 0)}~.
\eeq

The matrix element for $B(p) \to V_1(k_1,\epsilon_1) + V_2(k_2,\epsilon_2)$ can
be written
\beq
M = a \epsilon_1^* \cdot \epsilon_2^* + \frac{b}{m_B^2}(p \cdot
\epsilon_1)(p \cdot \epsilon_2) + i \frac{c}{m_B^2} \epsilon_{\mu \nu \rho
\sigma} p^\mu q^\nu \epsilon^{*\rho} \epsilon^{*\sigma}~;~~q \equiv k_1-k_2
\eeq
The transversity amplitudes depend on $a,b,c$ as $A_\parallel(a)$, $A_0(a,b)$, and
$A_\perp(c)$.  Under CP conjugation, $a \to \bar a,~b \to \bar b,~ i c \to -i
\bar c$.  Angular distributions depend on the angle $\phi$ and polar angles
$\theta_1$, $\theta_2$, each in the rest frame of the decaying $V_1$ or $V_2$:
$$
\frac{d \Gamma}{d \cos \theta_1 d \cos \theta_2 d \phi} \sim |A_0|^2 \cos^2
\theta_1 \cos^2 \theta_2 + (1/2)|A_\perp|^2 \sin^2 \theta_1 \sin^2 \theta_2
\sin^2 \phi
$$
$$
+(1/2) |A_\parallel|^2 \sin^2 \theta_1 \sin^2 \theta_2 \cos^2 \phi + (1/2\sqrt
{2}){\rm Re}(A_0 A^*_\parallel) \sin 2 \theta_1 \sin 2 \theta_2 \cos \phi
$$
\beq
- (1/2\sqrt{2}){\rm Im}(A_\perp A_0^*) \sin 2 \theta_1
\sin 2 \theta_2 \sin \phi - (1/2){\rm Im}(A_\perp A_\parallel^*) \sin^2
\theta_1 \sin^2 \theta_2 \sin 2 \phi~.
\eeq
The last two terms are T-odd and of two distinct types.

The interfering amplitudes are characterized by a weak phase difference
$\phi_w$ and a strong phase difference $\delta$.  In addition to the ``true''
TP ${\cal A}_T^{\rm true}$ defined above, one can define \cite{Datta} a
``fake'' TP:
\beq
{\cal A}_T^{\rm fake}
= \frac{\Gamma({\rm TP}
> 0) - \bar \Gamma({\rm TP} > 0) - \Gamma({\rm TP} < 0) + \bar \Gamma({\rm TP}
< 0)} {\Gamma({\rm TP} > 0) + \bar \Gamma({\rm TP} > 0) + \Gamma({\rm TP} < 0)
+ \bar \Gamma({\rm TP} < 0)}~,
\eeq
where TP$_{\rm true} \propto \sin \phi_w \cos \delta$,
TP$_{\rm fake} \propto \cos \phi_w \sin \delta$.  The two T-odd observables are
\beq
A_T^{(1)} \equiv \frac{{\rm Im}(A_\perp A^*_0)}
{|A_0|^2 + |A_\parallel|^2 + |A_\perp|^2}~,~~
A_T^{(2)} \equiv  \frac{{\rm Im}(A_\perp A^*_\parallel)}
{|A_0|^2 + |A_\parallel|^2 + |A_\perp|^2}~.
\eeq
For CP conjugates, one has similar definitions with barred amplitudes and
a minus sign from complex conjugation of the imaginary coefficient of $c$.
The TP asymmetries ${\cal A}_T$ then satisfy
\beq
{\cal A}_T^{\rm true} \propto {\rm Im}(A_\perp A_i^* - \bar A_\perp \bar
A_i^*)~,~~ {\cal A}_T^{\rm fake} \propto {\rm Im}(A_\perp A_i^* + \bar
A_\perp \bar A_i^*)~,~~(i=0,\parallel)~.
\eeq
The observables $A_T^{(1,2)}$ are related to those in Dorigo's talk
\cite{Dorigo} by
``$u$'' $\leftrightarrow A_T^{(2)}$; ``$v$'' $\leftrightarrow A_T^{(1)}$;
he reports on their measurement in $B_s \to \phi \phi$.

The decays $B \to \phi K^*$ and $B_s \to \phi \phi$ are both dominated by the
$b \to s$ penguin diagram.  Factorization predicts dominant longitudinal
polarization of the vector mesons, in contrast to observations
\cite{CDF10120,Ba99,Ba97} (Table \ref{tab:fact}).  By contrast, the
tree-dominated decay $B^0 \to \rho^+ \rho^-$ has $f_L= 0.992 \pm 0.024^{+0.026}
_{-0.013}$ \cite{Barr}, or nearly 1 as predicted.  There is no reason to trust
factorization for the penguin amplitude, which may be due to rescattering from
charm-anticharm intermediate states.

\begin{table}
\caption{Longitudinal and transverse fractions $f_L$ and $f_T$ for some
$b\to s$-penguin $B \to VV$ processes.
\label{tab:fact}}
\begin{center}
\begin{tabular}{c c c c c} \hline \hline
      & $B_s \to \phi \phi$ & $B^+ \to \phi K^{*+}$ & $B^+ \to \rho^0 K^{*+}$
 & $B^0 \to \rho^0 K^{*0}$ \\
 & \cite{CDF10120} & \cite{Ba99} & \cite{Ba97} & \cite{Ba97} \\ \hline
$f_L$ & 0.348$\pm$0.041$\pm$0.021 & 0.49$\pm$0.05$\pm$0.03 &
 0.52$\pm$0.10$\pm$0.04 & 0.57$\pm$0.09$\pm$0.08 \\
$f_T$ & 0.652$\pm$0.041$\pm$0.021 & 0.51$\pm$0.05$\pm$0.03 &
 0.48$\pm$0.10$\pm$0.04 & 0.43$\pm$0.09$\pm$0.08 \\ \hline \hline
\end{tabular}
\end{center}
\end{table}

From $B^0 \to \phi K^{*0}$ amplitudes quoted by \cite{Datta} we estimate
\beq
A_T^{(1)}=-0.260\pm0.048;~\bar A_T^{(1)}=0.203\pm0.050;~
 A_T^{(2)}=0.005\pm0.070;~ \bar A_T^{(2)}=0.010\pm0.064.
\eeq
These values imply a large fake $A_T^{(1)}$ (since $A_T^{(1)} - \bar A_T^{(1)}
\ne 0$); no true $A_T^{(1)}$ (since $A_T^{(1)} + \bar A_T^{(1)}$ is consistent
with zero); and no fake {\it or} true $A_T^{(2)}$ (since both $A_T^{(2)}$ and
$\bar A_T^{(2)}$ are consistent with zero).  The large fake $A_T^{(1)}$ simply
reflects the importance of strong final-state phases.

\section{$B_s \to J/\psi \phi$ vs.\ $B_s \to J/\psi f_0$}

Helicity or transversity analysis for $B_s \to J/\psi \phi$ (S-, P-, D-wave)
is avoided for $B_s \to J/\psi f_0$ (pure P-wave).  As CP($J/\psi$) = CP($f_0$)
= +, the overall final state is CP odd.  An estimate of the rate for this
process \cite{SZ} is
\beq
R_{f_0/\phi} \equiv \frac{\Gamma(B_s \to J/\psi f_0,~f_0 \to \pi^+
\pi^-)}{\Gamma(B_s \to J/\psi \phi,~\phi \to K^+ K^-)} \simeq 20\%~,
\eeq
to be compared with experimental values $0.252^{+0.046+0.027}_{-0.032-0.033}$
\cite{LHCbf}, $\simeq 0.18$ ($\sim 30\%$ stat.\ error) \cite{Bellef}, and
$ 0.292 \pm 0.020 \pm 0.017$ \cite{Dorigo}.  The CKM structure for this
process is the same as for $B_s \to J/\psi \phi$.  Although $f_0$ decays mainly
to $\pi \pi$, it seems to be ``fed'' mainly from $s \bar s$: Comparing $J/\psi
\to \phi \pi \pi$ and $J/\psi \to \omega \pi \pi$ \cite{FP77}, one
sees a $\pi \pi$ peak at $M(f_0) \simeq 980$ MeV in
$\phi \pi \pi$, not $\omega \pi \pi$.

\section{New physics constraints}

Two (of $\sim 100$) theoretical analyses \cite{Ligeti,Bai} emphasize the
correlation between $a_{sl}^q$, $\Delta m_q$, $\Delta \Gamma_q$, and the mixing
angle $\phi_q$, where $A^b_{sl} = (0.506 \pm 0.043) a^d_{sl}+(0.494 \pm 0.043)
a^s_{sl}$.  The questions of whether $\beta_s$ or $a_{sl}^q$ are nonstandard
are separate; they are related by $a_{sl}^q = (|\Delta \Gamma_q|/\Delta m_s)
\tan \phi_q$.  If the D0 dimuon asymmetry is mainly from $a^s_{sl}$, Ref.\
\cite{Bai} finds $a^s_{sl} = (-12.5 \pm 4.8) \times 10^{-3}$ by combining
with the D0 measurement $(-1.7 \pm 9.1) \times 10^{-3}$.  Using in this
formula the (CDF, LHCb) average $\Delta m_s = (17.70 \pm 0.08)$ ps$^{-1}$ and
the (CDF, D0) average $\Delta \Gamma_s=0.094 \pm 0.031$ ps$^{-1}$, one expects
$\phi_s = (-67^{+18}_{-7})^\circ$.  Comparing with $\phi^s_M = (-39 \pm 17)^
\circ$, this would favor slightly larger $\Delta \Gamma_s$ or a nonstandard
value of $a^d_{sl}$.  In Ref.\ \cite{LNCKMf} it is noted that one must respect
the SM prediction of $\Delta m_q$.  New physics must affect mainly {\it phases}
of mixing amplitudes.

\section{A cursory look at new physics scenarios}

Supersymmetry has generic flavor-changing (but controllable) effects
\cite{SUSY}.  Randall-Sundrum \cite{RS} scenarios in which different quarks
lie at
different points along a fifth dimension offer a language for understanding
quark mixings; but there is no predictive scheme yet.  Theories with an extra
(flavor-changing) $Z$ can induce mixing as desired.  In Ref.\ \cite{Bai}
a contribution to $\Delta \Gamma$ is introduced through a new light
pseudoscalar (an on-shell state in $B_s \leftrightarrow \bar B_s$).  These are
just some examples of a wealth of models on the market.  Some
of them predict other observable consequences but there are too many to
enumerate exhaustively.  Two of my current favorites are (1) a fourth
generation, and (2) a hidden sector.

Lunghi and Soni \cite{LS} note the tension between $\sin 2 \beta = \sin 2
\phi_3 = 0.668 \pm 0.023$ (measured in $B$ decays) and that ($0.867 \pm 0.048$)
in (their) CKM fit.  They note effects of new physics on both $\Delta
{\rm Flavor}=1$ (penguin) and $\Delta {\rm Flavor}=2$ (box) amplitudes but
give no specifics on $\beta_s$ or $a_{sl}^s$.

\begin{figure}[h]
\begin{center}
\includegraphics[width=0.75\textwidth]{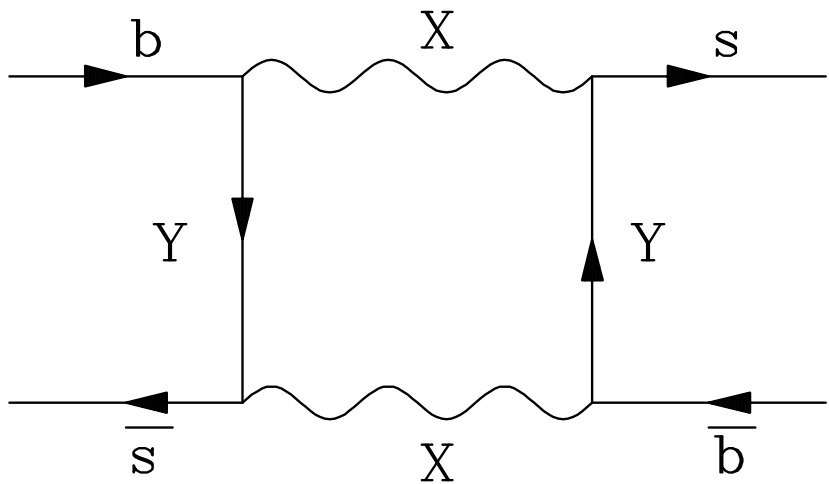}
\end{center}
\caption{Diagram utilizing a hidden sector describing $B_s$--$\ob_s$ mixing.
\label{fig:xybox}}
\end{figure}

In a ``hidden sector'' let an extended gauge sector $G$ describe dark matter,
and let there be particles $Y$ with charges in both the SM and in $G$, and
particles $X$ with charges only in $G$.  A box diagram describing
$B_s$--$\ob_s$ mixing in this scenario is shown in Fig.\ \ref{fig:xybox}.
Table \ref{tab:types} gives examples of ordinary, mixed, and ``shadow'' matter.
There are clearly many opportunities in such a scenario for new contributions
to penguin and box diagrams.

\begin{table}[h]
\caption{Types of matter and their SM and hidden charges.
\label{tab:types}}
\begin{center}
\begin{tabular}{c c c c} \hline \hline
Type of matter & Std.\ Model &    G    & Example(s) \\ \hline
Ordinary       & Charged & Uncharged & Quarks, leptons \\
Mixed ($Y$)    & Charged & Charged & Superpartners \\
Shadow ($X$)   & Uncharged & Charged & $E_8'$ of E$_8 \otimes$
E$_8'$ \\ \hline \hline
\end{tabular}
\end{center}
\end{table}

\section{Summary}

$B_s$ decays and mixing provide potential mirrors of new physics.  While the
phase $\beta_s$ has moved toward its Standard Model value, even the currently
measured value of $\beta_s$ should be manifested in time-dependent quantities.

The D0 collaboration \cite{D0mumu} claims a dimuon charge asymmetry.
At this conference \cite{Harr} CDF has reported a remeasurement of $\bar \chi$
and we look forward to their further progress on dimuons.  The signal requires
subtraction of a big kaon background.  Is what's left really due to $b$ quark
decays?  We have proposed an impact parameter cut of $b < 100 ~\mu$m to find
out \cite{GRmumu}.

Using triple products in four-body decays, one can construct T-odd observables
providing strong and weak phase information.  There is interest in what new
physics one can learn from $B_s \to \phi \phi$ \cite{Dorigo}.

As for whether there is new physics in any of the above hints, I urge you to
have your favorite model ready; there are enough to go around.

\section*{Acknowledgments}

We are grateful to B. Bhattacharya,
M. Dorigo, D. London, and D. Tonelli for helpful
discussions.  The work of J.L.R. was supported in part by the United States
Department of Energy, Grant No.\ DE-FG02-90ER40560.

\end{document}